\documentclass[prd, twocolumn, nofootinbib, floatfix]{revtex4-1}

\usepackage{amsmath}
\usepackage{graphicx}
\usepackage{dcolumn}
\usepackage{bm}
\usepackage{epsfig}
\usepackage{amssymb,latexsym,mathrsfs}
\usepackage{graphicx}
\usepackage{color}
\usepackage{hyperref}

\hypersetup{
    colorlinks=true,
    linkcolor=red,
    citecolor=blue,
} 

\newcommand{\be}{\begin{equation}}
\newcommand{\ee}{\end{equation}}
\newcommand{\bs}{\begin{split}} 
\newcommand{\bea}{\begin{eqnarray}}
\newcommand{\eea}{\end{eqnarray}}

\newcommand{\om}{\Omega_m}
\newcommand{\omp}{\Omega_{m,0}}

\newcommand{\geff}{G_{\rm eff}} 
 
\newcommand{\gm}{\gamma} 
 
\newcommand{\dl}{\delta}

\newcommand{\lcdm}{$\Lambda$CDM}

\begin{document}

\title{The End of Cosmic Growth} 

\author{Eric V.\ Linder${}^{1,2}$, David Polarski${}^{3}$} 
\affiliation{${}^1$Berkeley Center for Cosmological Physics \& Berkeley Lab, 
University of California, Berkeley, CA 94720, USA\\ 
${}^2$Energetic Cosmos Laboratory, Nazarbayev University, 
Astana, Kazakhstan 010000\\ 
${}^3$Laboratoire Charles Coulomb, Universit\'e de Montpellier \& CNRS UMR 5221, F-34095 Montpellier, France}

\begin{abstract} 
The growth of large scale structure is a battle between gravitational attraction and cosmic acceleration. 
We investigate the future behavior of cosmic growth under both general relativity (GR) and modified gravity during prolonged acceleration, deriving analytic asymptotic 
behaviors and showing that gravity generally loses and growth ends. We also note the ``why now'' problem 
is equally striking when viewed in terms of the shut down of growth.  
For many models inside GR the gravitational growth index $\gamma$ also shows today as a unique time between constant behavior in the past and a higher asymptotic value in the future.  
Interestingly, while $f(R)$ models depart in this respect dramatically from GR today and in the recent past, their growth indices are identical in the asymptotic future and past.  
 
\end{abstract}

\date{\today} 

\maketitle

\section{Introduction} 
The growth of cosmic structure is what allows us to exist, taking initial seeds of density inhomogeneity from quantum fluctuations in inflation and under gravitational instability forming the massive structures of galaxies and clusters of galaxies. Yet we also live in a presently accelerating universe, and this is already shutting down cosmic growth. 

We explore the future consequences of the battle between gravitational attraction (possibly even stronger than in GR) and the 
accelerating expansion, whether due to dark energy (with Newton's constant unaffected) or modified gravity. Our focus is on the future and we derive the conditions for the end of cosmic growth, in terms of the growth rate and the gravitational growth index $\gm$. 
While \cite{0908.2669} explored possible values of $\gamma$ today and in the recent past as well as their model and scale dependence, \cite{1610.00363} addressed also the future asymptotic behaviour $\gamma_{\infty}$ however only inside GR. 
Early papers such as \cite{P84,fry,lights,LLPR91,wangs,amen,dhw} focused mostly near the present, and did not use $\gm$ as a test of growth separate from the expansion effects until \cite{groexp}. 

A linear density perturbation $\dl\equiv\dl\rho/\rho$ evolves with scale factor $a$ according to its growth factor $D(a)$ as $\dl(a)/\dl(a_i)=D(a)/D(a_i)$. The cosmic growth rate 
$f\equiv d\ln D/d\ln a$ 
is especially useful to characterize its slow down, and for a given Fourier mode $f$ evolves in subhorizon, linear theory according to 
\be 
f'+f^2+\left[2+\frac{1}{2}\frac{d\ln H^2}{d\ln a}\right]\,f-\frac{3}{2}\geff(a,k)\om(a)=0\ , 
\label{eq:ffull} 
\ee 
where a prime denotes the derivative  
$d/d\ln a$, $H$ is the Hubble expansion rate, $\om(a)$ is the matter density as a fraction of the critical density, and $\geff(a,k)$ is the (dimensionless) effective time and scale dependent gravitational strength 
divided by Newton's constant $G$ (see e.g.\ \cite{BEPS00}). 
We can write the growth rate as $f=\om^\gm(a)$, 
which defines the growth index $\gm=\ln f/\ln\om(a)$.

\section{Future Growth in General Relativity} \label{sec:gr} 
\begin{figure}[htbp!]
\includegraphics[width=0.9\columnwidth]{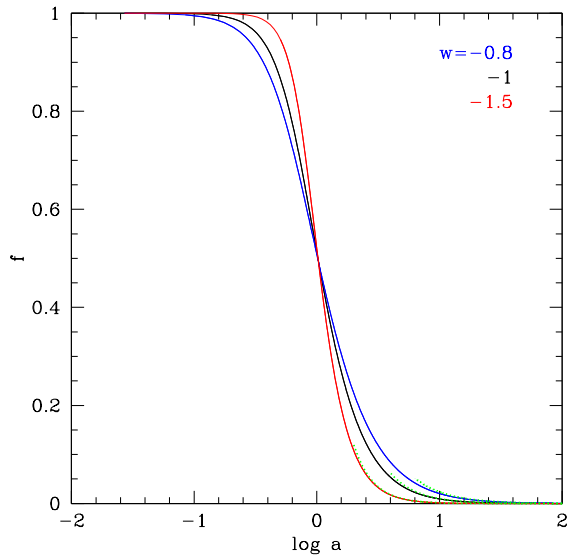}
\caption{
The growth rate $f$ shows steep behavior vs $\log a$, 
even for models far from \lcdm\ ($w=-1$). 
Growth is suppressed relative to the matter dominated era  ($f(a\ll1)=1$) 
as cosmic acceleration begins near today ($a=1$), and undergoes a 
sharp transition shutting off growth. The three solid curves show the behavior for different values of the effective dark energy equation of state $w$. The short green dotted curves at $a\gg1$ give the asymptotic behavior $f\propto a^{(3w-1)/2}$ for each curve. 
} 
\label{fig:flga} 
\end{figure}

Taking gravity 
to be described by general relativity (GR), $\geff=1$. 
Cosmic acceleration enters growth through the Hubble friction term and the diminished source term $\om(a)$. It has a rather dramatic impact on the growth rate. We see in Fig.~\ref{fig:flga} 
that $f$ plunges from 90\% of its matter dominated value ($f=1$) to 10\% in less 
than 2 e-folds of expansion. The present value is close to the middle of this sharp cutoff  ($f_{\Lambda{\rm CDM}}\approx 0.5$ for the present value of the matter density fraction $\omp\approx0.3$). One can view this as an alternate, growth view of the coincidence or ``why now'' problem familiar from the expansion history. 
This reflects in the departure of $\gm$ starting around the present time from a nearly constant behavior in the past \cite{1610.00363,groexp}.

We can analytically derive the late time asymptotic behavior of the suppression as 
\be 
f \sim c_f\,a^{(3w-1)/2} \ . \label{fasymp}
\ee 
As the growth rate goes to zero, the density perturbation freezes, 
$D\to D_\infty$ and cosmic growth ends. 
  
Within GR, the expansion history fully determines the growth history (in the linear regime, with negligible perturbations 
in components other than matter, and given the initial conditions). 
To form a test of general relativity, \cite{groexp} 
separated out the effects on the growth of the cosmic expansion from the gravitational coupling strength using the growth index $\gm$. 
For observational data (i.e.\ at $a\le1$) cosmic growth can be accurately described in many cosmologies by a constant value for $\gm$ \cite{groexp,lincahn}. For example, for smooth noninteracting models including $\Lambda$CDM, within GR, the growth amplitude $D(a)$ is fit to within 0.1\% of the exact value by using $\gm=0.55$ and the growth rate $f(a)$ to within 0.3\%. Note that next generation data is expected to constrain these quantities at about the percent level, so this approximation is sufficient for a consistency test of such models. 

\begin{figure}[htbp!]
\includegraphics[width=\columnwidth]{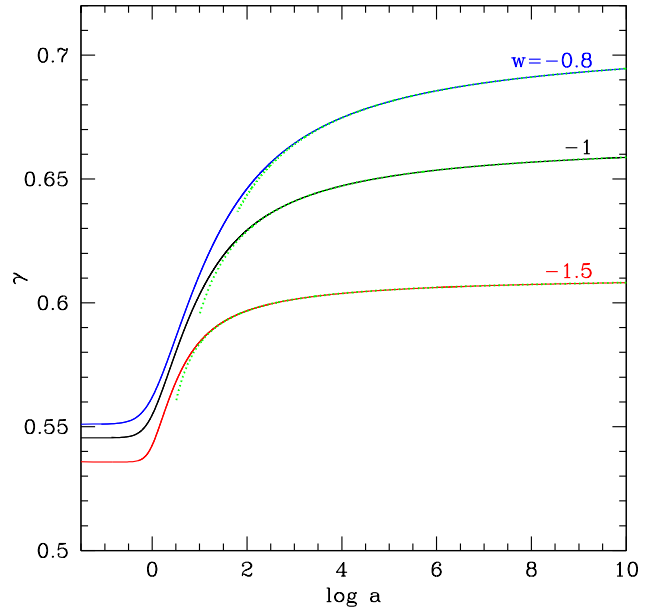}
\caption{
The gravitational growth index $\gm$ shows sudden evolution in the near future, after a predominantly constant behavior in the past. 
Although growth freezes in the future, $\gm$ asymptotically goes to a new finite constant value because $\Omega_m\to 0$ too. 
The three solid curves show the behavior for different values of $w$. 
The green dotted curves at $a\gg1$ give the asymptotic behavior $\gm=\gm_\infty+c_\gm/[\ln\om(a)]$ for each curve. 
} 
\label{fig:gamlnax} 
\end{figure}

However, the near constancy of $\gm$ until today \cite{PG07} is due to the relatively recent onset of cosmic acceleration. We find a very different behavior for future growth. The growth index rapidly rises starting near the present, indicating that the growth rate $f$ is more sensitive to the diminishing matter density fraction $\om(a)$ and hence weakens rapidly. However, $\gm$ then 
tends to a new higher asymptotic value $\gm_\infty$. The approach goes inversely with the logarithm of the matter density \cite{1610.00363}, 
\be 
\gm(a\to \infty)\sim \frac{3w-1}{6w} + \frac{c_\gm}{\ln\om(a)}\equiv                 \gm_\infty+\frac{c_\gm}{\ln\om(a)} \ . 
\ee 
Note that since at late times $\om(a)\approx [\omp/(1-\omp)]a^{3w}$ then $\ln\om(a)\approx 3w\ln a$ and $\gamma_\infty$ is recovered using Eq.~(\ref{fasymp}). For example, within general relativity and $\Lambda$CDM, $\gm_\infty=2/3$. 
For arbitrary $w$ we can just use the asymptotic value $w_{\infty}$. 
Figures~\ref{fig:flga} and \ref{fig:gamlnax} illustrate these results and Table~\ref{tab:gamf} summarizes the late time asymptotic behaviors for 
three different values of the effective dark energy equation of state.

\begin{table} 
\begin{center} 
\begin{tabular}{l|c|c|c|c} 
Model & \ $\gm_\infty$ \ & \ $c_\gm$ \ & \ $\left(d\ln f/d\ln a)\right)_\infty$ \ & \ $c_f$ \ \\ 
\hline  
\rule{0pt}{1.1\normalbaselineskip}$w=-1$ & $2/3$ & $0.553$ & $-2$  & \ 0.989 \ \\ 
$w=-0.8$ & \ $0.708$ \ & \ $0.772$ \ & $-1.7$  & \ 1.19 \ \\ 
$w=-1.5$\ \ \ & $0.611$ & $0.309$ & $-2.75$ & \ 0.811 \ \\ 
\end{tabular}
\end{center}
\caption{Values for the constants entering the asymptotic formulas for the 
gravitational growth index $\gm$ and the growth rate $f$. While $\gm_\infty$ and 
$d\ln f_\infty/d\ln a$ can be derived analytically, the coefficients 
$c_\gm$ and $c_f$ depend on the entire growth evolution and are found numerically. 
} 
\label{tab:gamf} 
\end{table} 

\section{Future Growth in Modified Gravity} \label{sec:modgr} 
Modified gravity enters through the source term in the growth equation, as shown by the factor $\geff$ in Eq.~(\ref{eq:ffull}). This loosens the connection between 
expansion history and growth history, and has the 
potential to change the balance between the cosmic 
growth battle. Interestingly, 
with regard to asymptotic future growth, 
we find that if the source term involving $\geff(a)\om(a)\ll f$ then 
$\geff$ will not affect the asymptotic behavior $f\sim  a^{(3w-1)/2}$  derived in the previous section. Thus 
growth still loses against acceleration. 

How strong does gravity need to become to allow growth to continue? 
We need that $\geff(a)\om(a)$ does not tend to zero. 
Since we are interested in asymptotic behavior, consider  $\geff\propto a^p$ asymptotically. This then gives the condition  
$p+3w\ge 0$ ($p\ge 3$ for 
\lcdm). When $p+3w < 0$ the growth will eventually end, $f\to0$, however the approach to 
shutdown can differ depending how gravity strengthens at late times (just being stronger today is insufficient). 
Two cases can arise for $p<-3w$: looking at $\geff(a)\om(a)\sim a^{p+3w}$ against $f_{\rm GR}\sim a^{(3w-1)/2}$, 
we see that the dividing behavior occurs for 
$p=(-3w-1)/2$ (so $p=1$ for a \lcdm\ background). For smaller $p$, 
i.e.\ gravity strengthening but not fast enough, the asymptotic behaviors of $f$ and $\gm$ are the same as GR. For larger $p<-3w$, growth will still end but with a different asymptotic behavior. 
All the behaviors are summarized in Table~\ref{tab:tgeff}.

\begin{table} 
\begin{center} 
\begin{tabular}{l|c|c|c} 
$\geff\sim a^p$ & \ $f_\infty$ \ &  \ $\left(d\ln f/d\ln a\right)_\infty$ \ & \ $\gm_\infty$ \ \\ 
\hline  
\rule{0pt}{1.1\normalbaselineskip}$p\le\frac{-3w-1}{2}$ & 0 & 
$\frac{3w-1}{2}$ & $\frac{3w-1}{6w}$ \\ 
\rule{0pt}{1.05\normalbaselineskip}$\frac{-3w-1}{2}\le p< -3w$\ & 0 & $p+3w$ & $\frac{p+3w}{3w}$ \\
\rule{0pt}{1.05\normalbaselineskip}$p=-3w$ & const (Eq.~\ref{eq:fconst}) & 0 & 0 \\ 
\rule{0pt}{1.05\normalbaselineskip}$p>-3w$ & $\infty$ (Eq.~\ref{eq:fdiv}) & $p/2$ & $<0$ \\ 
\end{tabular}
\end{center}
\caption{Asymptotic growth behaviors are given for modified gravity, depending on how 
rapidly it strengthens asymptotically. 
} 
\label{tab:tgeff} 
\end{table}

The first case applies to GR ($p=0$) obviously, to modified gravity with weakening strength  
($p<0$), or even to gravity with strength increasing  
but not fast enough. 
For the second case the source term 
becomes larger than the perturbation decaying mode in the $\delta$ growth 
factor equation, 
or of the same order as the friction term in the $f$ growth rate equation.  
This is sufficient to alter $\gm_\infty$ and the asymptotic approach of $f\to0$, but growth still ends. 
Only when $p=-3w$ does growth continue. Then the source term stays 
constant and from Eq.~(\ref{eq:ffull}) $f\to\,$const 
with solution 
\be 
f=\frac{3w-1}{4}+\sqrt{\left(\frac{1-3w}{4}\right)^2+\frac{3}{2}\,\frac{\omp}{1-\omp}\,c_g} \ , \label{eq:fconst} 
\ee 
where $\geff(a)\sim c_g a^p$ asymptotically with $p=-3w$. 

Finally, an even more increasing gravitational 
strength with $p>-3w$ will cause the growth rate itself to increase. Then the $f^2$ term 
dominates over the friction term,  
giving the asymptotic solution 
\be 
f(a)\sim \sqrt{\frac{3}{2}\geff(a)\,\om(a)}\sim a^{p/2} \quad {\rm if}\ p>-3w \ . \label{eq:fdiv} 
\ee 
Strong enough gravity, i.e.\ $\geff(a)\sim a^{p\,>-3w}$, can overcome cosmic acceleration and 
continue the growth of cosmic structure in the future. However, this is a formal solution only since such increasing growth, with 
$D\sim e^{a^{p/2}}$, will rapidly invalidate the linear theory formalism used. 
We do not propose any specific theory of gravity with such properties, but note that the Horndeski class of general 
scalar-tensor theories have considerable freedom to enable high values of $p$. 

\begin{figure}[htbp!]
\includegraphics[width=\columnwidth]{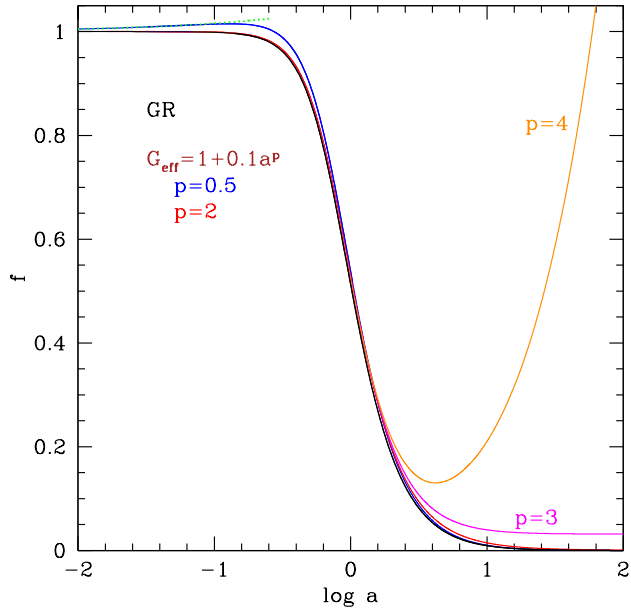}
\caption{
Modified gravity does not generally qualitatively alter the growth rate, its future 
vanishing, and today marking roughly the halfway point in the process. Only if the 
gravitational strength increases very rapidly, $\geff\sim a^p$ with $p\ge-3w$, is there a significant 
effect. The curves show $f$ for various values $p$, with $w=-1$ 
(\lcdm\ background). 
The green dotted curve at $a\ll1$ gives the early asymptotic behavior 
due to initial conditions different from general relativity (GR). 
} 
\label{fig:fgeff} 
\end{figure}

\begin{figure}[htbp!] 
\includegraphics[width=\columnwidth]{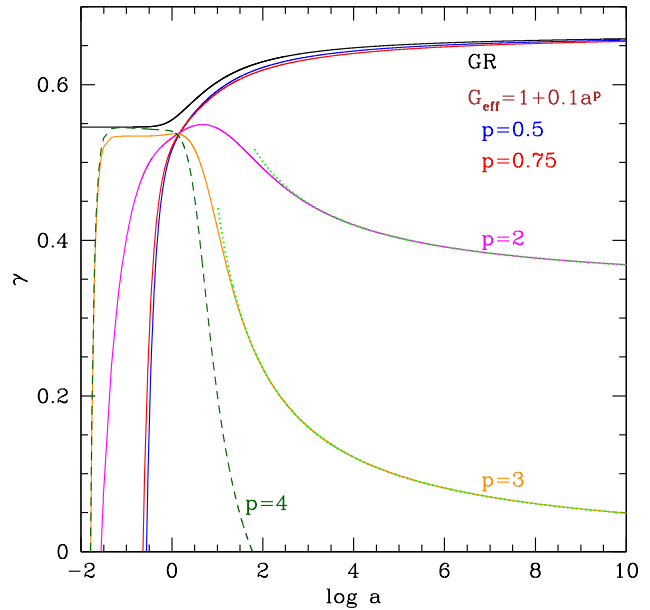}
\caption{
Modified gravity alters the gravitational growth index $\gm$ 
behavior at late times only if the gravitational strength increases rapidly enough. For $\geff\sim a^p$, the asymptote remains that of GR for $p<(-3w-1)/2$. The curves show $\gm$ for various values $p$, with $w=-1$ (\lcdm\ background). 
The green dotted curves at $a\gg1$ give the asymptotic late behavior 
for the $p=2$, 3 curves. 
Since enhanced gravity strengthens growth at early times, driving 
$f>1$, then $\gm<0$ at early times. 
} 
\label{fig:geffgam} 
\end{figure}

We verify all these behaviors numerically and illustrate them in Figs.~\ref{fig:fgeff} and \ref{fig:geffgam}. Adopting the form $\geff=1+c_g a^p$ we see that 
indeed when $p<(-3w-1)/2$ then the asymptotic growth has the same value of 
$\gm$ as for GR. (The asymptotic behavior is independent of 
$c_g$, which here we choose to be 0.1, simply for visual convenience.) When $p$ increases further 
(and $p<-3w$), the growth rate is still suppressed, despite the strengthening gravity, with an asymptotic gravitational growth index given by 
\be 
\gm=\frac{p+3w}{3w}=\gm_{\rm GR}-\frac{2p+3w+1}{6|w|}\ . 
\ee 
Increasing $p$ reduces $\gm$, which is a 
sign of increased relative growth (the growth rate itself may still be 
diminishing, just at a slower pace than in GR). 
The 
growth rate stops going to zero for $p=-3w$, where $\gm_{\infty}=0$, and so cosmic growth no longer 
ends for such increasing gravity; rather the growth factor 
$D\sim a^f$ where $f$ is the small constant value given by Eq.~(\ref{eq:fconst}). This is in contrast to all smaller $p$ where the growth amplitude $D$ freezes 
asymptotically -- the end of cosmic growth. 
Even stronger gravity, i.e.\ even higher $p$, drives $\gm$ negative and the growth 
rate climbs again, breaking the linear regime. 

Note that in some theories of modified gravity, a 
de Sitter asymptotic expansion implies that $\geff$ will freeze in the future (for 
example in $f(R)$ gravity the scalaron mass goes to a constant, see below), and so 
we expect the asymptote $p=0$, despite a present strengthening or weakening. Thus in such cases there is an end to cosmic growth. 
Also note that while our focus is on growth at late times, at early times an enhanced 
gravitational strength increases growth, leading to a growth rate $f>1$ and so, by 
$\gm=\ln f/\ln\om$, a negative $\gm$. This is discussed in \cite{1302.4754}, and we 
exhibit the analytic formula derived there, $f=1+[3/(5+2p)]c_g a^p$ 
in Fig.~\ref{fig:geffgam} as an excellent match to the numerical solution.

\begin{figure}[htbp!]
\includegraphics[width=\columnwidth]{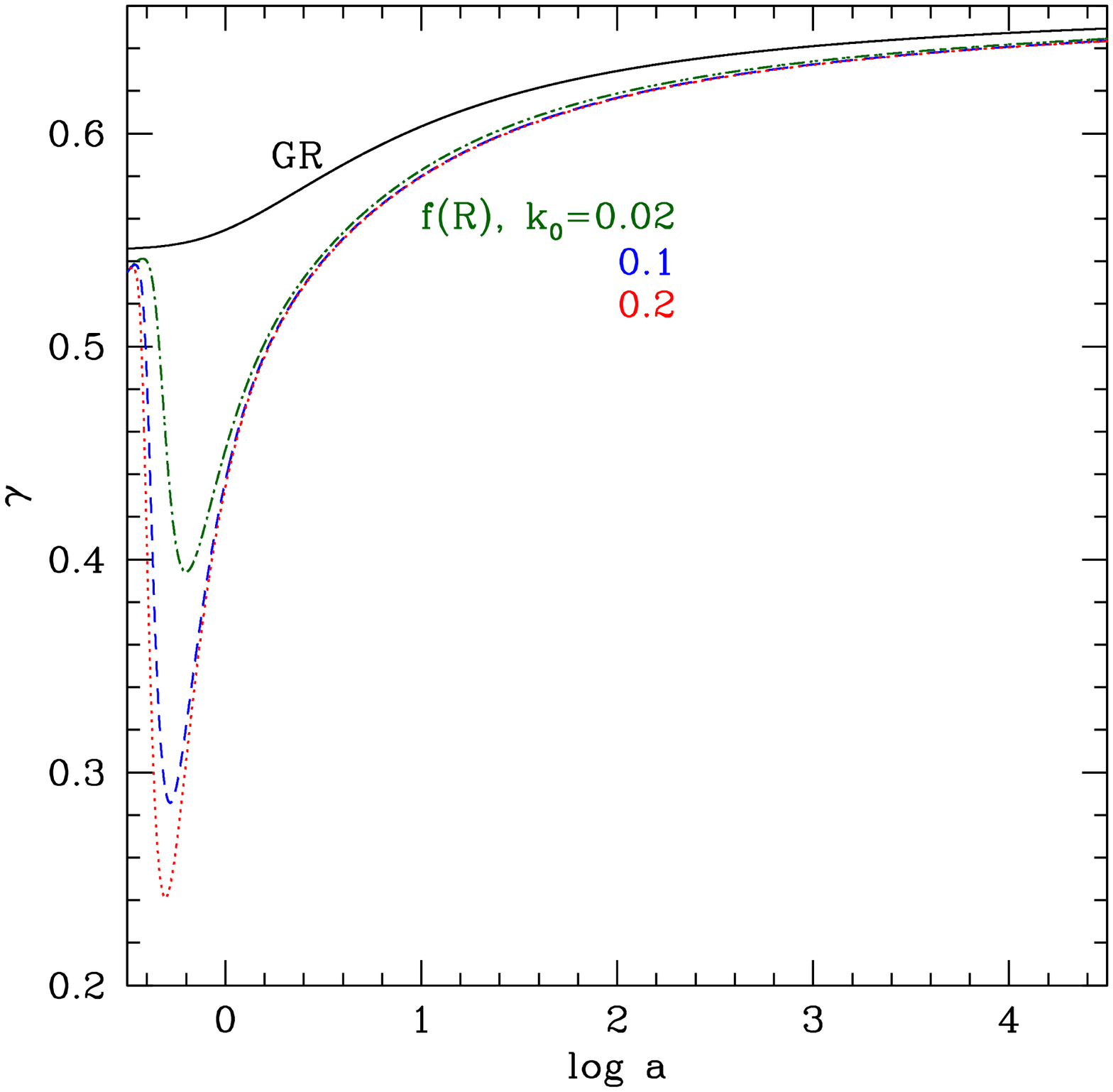}
\caption{
$f(R)$ gravity gives scale dependent growth, but its asymptotic future behavior is similar to that of GR 
and the scale dependence vanishes as $k/(aM)\to0$ for large $a$. Curves are labeled with $k_0$, the density mode wavenumber today, in units of $h$/Mpc. Near the present, growth is enhanced, with $\gm\approx0.42$ \cite{GMP08}, $(d\gamma/da)_0\approx 0.25$, and a larger growth rate $f$ than GR. Note that $\gm\le \gm_{\rm GR}$.  
} 
\label{fig:frgam} 
\end{figure}

\begin{figure}[htbp!]
\includegraphics[width=\columnwidth]{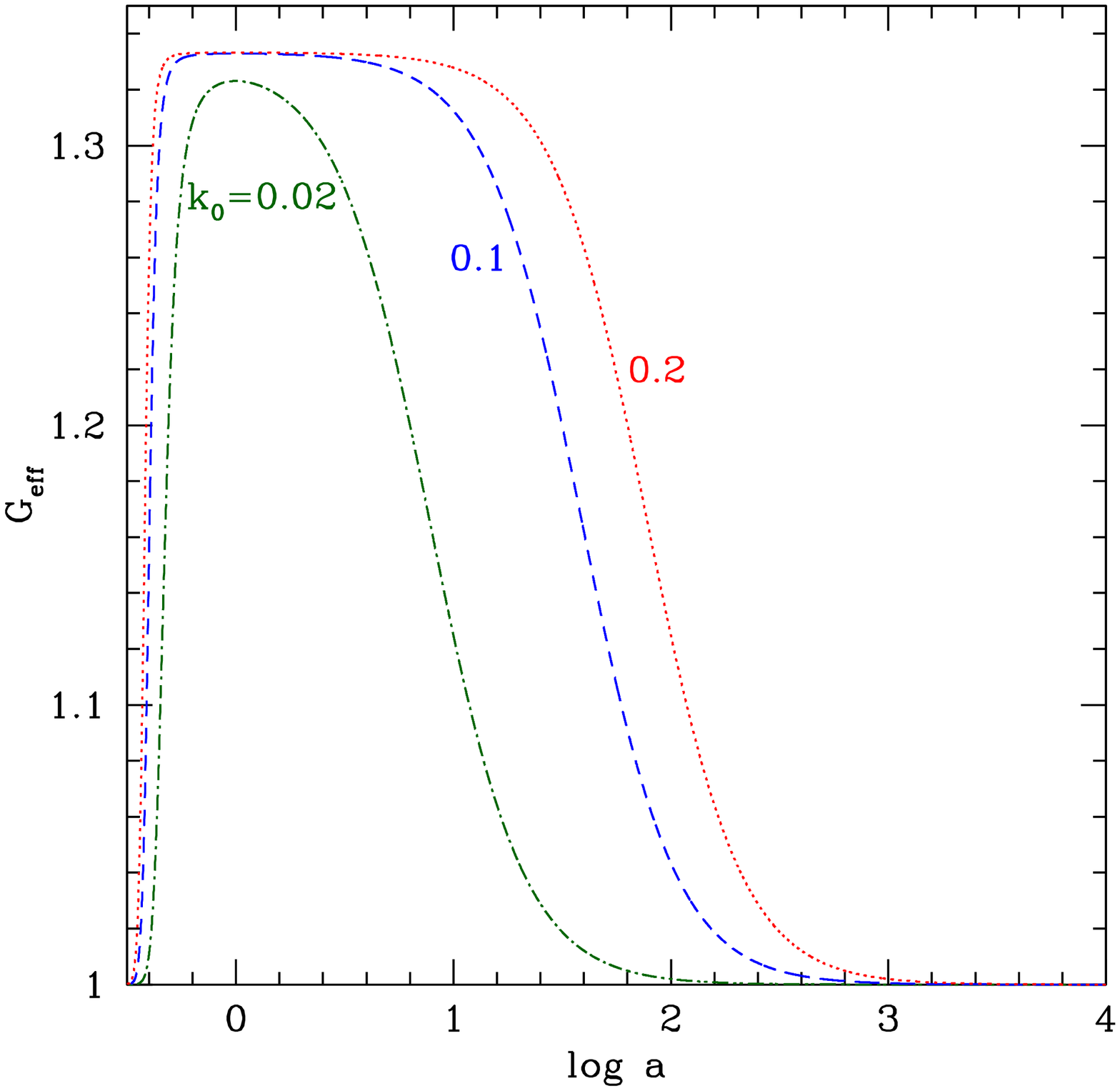}
\caption{
$f(R)$ gravity has enhanced, scale dependent gravitational strength with a maximum at 
$(4/3)G$, but that approaches GR at early and 
late times.} 
\label{fig:frgeff} 
\end{figure}

Finally, as an example with scale dependent growth, we investigate an exact $f(R)$ gravity model. These models are important representatives of viable modified gravity models.  We use the exponential gravity model of 
\cite{expfr}.  
In the approach to the late time de Sitter state the Ricci scalar and scalaron mass freeze. Thus the effective gravitational strength, which can rise to $4/3$ times Newton's constant near the present for scales small enough, 
goes back to its GR value 
in the future. 
This corresponds to the $p=0$ asymptote and hence 
cosmic growth indeed ends and with asymptotic behavior similar to that of GR. 
Figure~\ref{fig:frgam} exhibits this $f(R)$ model's exact numerical solution for the scale dependent growth and Fig.~\ref{fig:frgeff} shows the numerical evolution of the 
gravitational strength, illustrating the freezing to GR.

There are worse things than the end of cosmic growth -- growth can actually be 
undone in a Big Rip scenario with $w<-1$ \cite{bigrip}. However that breaks 
assumptions used here such as the subhorizon approximation. Conversely, if the dark 
energy itself can cluster on subhorizon scales, then the growth equation gains 
an additional source term. For the great majority of models as considered here, 
though, cosmic growth ends.

\section{Conclusions} \label{sec:concl} 
In general there is an end to cosmic growth and today is unique in that roughly an efold ago the growth rate was greater than 90\% of its maximum, and roughly an efold in the future it will be 90\% of the way to vanishing. 
While accelerating universes in GR lead to an end of cosmic growth, the asymptotic growth can be discriminated through the growth index $\gamma$ as it is affected by the perturbation decaying mode. 

Even in modified gravity with strengthened gravity, 
acceleration generally wins and growth ends. We detail the conditions under which this -- or alternate outcomes -- occurs. 
Interestingly, $f(R)$ gravity and \lcdm\ GR have identical growth behaviors in both asymptotic future and past. 
At the present time however, they differ strongly with for $f(R)$ a dip in the growth index in the recent past resulting from a bump in $G_{\rm eff}$, a substantially lower value today $\gamma_0\approx 0.42$, and a large slope.  
We happen to live at the right epoch in order to detect 
a decisive observational signature of these models.

\acknowledgments 

We thank Arman Shafieloo and KASI for hospitality during (and for DP for the month following) the 5th Korea-Japan Workshop 
on Dark Energy, where this project started. EL is supported in part by the Energetic Cosmos Laboratory and by the U.S.\ Department of Energy, Office of Science, Office of High Energy Physics, under Award DE-SC-0007867 and contract no.\ DE-AC02- 05CH11231.


\end{document}